\documentstyle[aps,pra,multicol]{revtex}
\begin{document}
\title{Maximum thickness of a two-dimensional trapped Bose system}
\author{Sang-Hoon Kim}
\address{Division of Liberal Arts, Mokpo National Maritime University,
 Mokpo 530-729, Korea}
\maketitle
\begin{abstract}
The trapped Bose system can be regarded as two-dimensional
 if the thermal fluctuation energy is less than the lowest energy 
in the perpendicular direction.
Under this assumption, we derive an expression for the maximum 
thickness of an effective two-dimensional trapped Bose system.
\end{abstract}
\draft
\begin{multicols}{2}

	 It has been known that the Bose-Einstein Condensation (BEC)
 cannot occur in either two-dimensional (2D)
 or one-dimensional (1D) uniform Bose gas at a finite temperature
 because thermal fluctuations destabilize the condensate \cite{hohe}.
	However, when a spatially varying potential which breaks
 the translational invariance exists, the
BEC may occur in low dimensional inhomogeneous systems.
In the presence of  harmonic trapping, the effect of thermal fluctuations
are strongly quenched due to the different behavior exhibited 
by the density of states.  

In three-dimensional (3D) traps, the experimental results for the BEC 
have been obtained assuming that the thermal fluctuation energy $k_BT$ 
 is much larger than all the  oscillator energies 
($\hbar\omega_x, \hbar\omega_y,\hbar\omega_z$).
In order to achieve a 2D BEC in the trap, it is necessary to choose
 the frequency $\omega_z$ large enough to satisfy the condition
$\hbar \omega \ll k_B T_{2D} \ll \hbar \omega_z$, 
where $\omega = \sqrt{\omega_x \omega_y}$ and  $T_{2D}$
is the 2D transition temperature.
This is a rather difficult condition to satisfy in the trap design,
and leads the realization of a 2D system in another way.

Recently, Safonov {\it et al.}   \cite{safo} have reported an observation of a
quasi-2D BEC  in liquid hydrogen layers.
They successfully confined hydrogen atoms on a liquid $^4$He surface 
which corresponds to a potential well of  20 $\mu$m width.
Also, Gauck {\it et al.} \cite{gauc} claimed that they achieved another
 quasi-2D system of argon atoms  confined in a planer matter waveguide 
in the close vicinity of  $\mu$m.
However, the question that up to what thickness we can regard
 the system as 2D is remained.
The mathematical concept of 2D which neglects one (z-direction)
degree of freedom and allows particles to move only in surface (x-y plane)
is not physically acceptable due to the uncertainty principle.
In this communication, we suggest a criterion for the BEC to exhibit 
2D behavior in 3D space, and obtain a maximum thickness of the 
2D trapped Bose system.

In the 2D experimental setup, the z-directional thickness corresponds to 
 ideal  rigid walls.
For the infinite  potential well of $-d/2 \le z \le d/2$,
the lowest energy is given by
$E_g = \hbar^2 \pi^2 / 2 m d^2$.
The system can be regarded as 2D as long  as the thermal fluctuation
energy is less than the z-directional lowest energy.  That is,
\begin{equation}
 k_B T_{2D} <  \frac{\hbar^2 \pi^2 }{ 2 m d^2}.
\label{30}
\end{equation}

The transition temperature of the BEC is not precisely known except for
the ideal Bose gas in the harmonic trap.  Although the systems used 
in the experiment were not ideal, the measured transition temperatures
 in 3D were found to be very close to the ideal gas value.
A similar situation is expected in 2D.
The transition temperature of the 2D BEC for the ideal Bose gas system
in the harmonic trap is given by  \cite{mull}
\begin{equation}
k_B T_{2D} = \hbar \omega \left(\frac{N}{\zeta(2)} \right)^{1/2},
\label{10}
\end{equation}
where $N$ is the number of atoms in the trap and
 $\zeta(x)$ is the Riemann-Zeta function.

Substituting  Eq. (\ref{10}) into Eq. (\ref{30}),
 we obtain an effective 2D  thickness of
\begin{equation}
\frac{d}{a_{ho}} < \frac{\pi}{\sqrt{2}}
 \left( \frac{\zeta(2)}{N} \right)^{1/4},
\label{40}
\end{equation}
or
\begin{equation}
{\rm (Thickness\, of\, 2D)} \, < \,  \frac{2.516}{N^{1/4}}\; a_{ho},
\label{50}
\end{equation}
where $a_{ho}$ is the harmonic oscillator length given as
$a_{ho} = \sqrt{\hbar/m\omega}$.
We note that a typical value for $a_{ho}$ is of the order of
 $\mu$m for alkali Bose atoms and much larger for hydrogen atoms.

In conclusion, we have obtained a maximum value of the effective 
thickness of a 2D trapped Bose system in which to observe the BEC.
The maximum radius of a 1D trapped Bose system could be obtained
in a similar way.

We thank Professors C.K. Kim, K. Nahm, and M. Chung for useful discussions.


\end{multicols}
\end{document}